\documentclass{article} 
\usepackage{multirow}

\usepackage{iclr2020_conference,times}

\usepackage{placeins}

\usepackage{subfigure}

\usepackage{amsmath,amsfonts,bm}

\def\eqref#1{equation~\ref{#1}}

\def\1{\bm{1}}

\DeclareMathAlphabet{\mathsfit}{\encodingdefault}{\sfdefault}{m}{sl}
\SetMathAlphabet{\mathsfit}{bold}{\encodingdefault}{\sfdefault}{bx}{n}

\usepackage{hyperref}

\numberwithin{table}{section}

\usepackage{url}

\usepackage{graphicx}
\graphicspath{{Figs/}}
\DeclareGraphicsExtensions{.eps,.pdf,.jpg,.gif}

\title{Automated Smartphone based System for \\Diagnosis of Diabetic Retinopathy}

\author{Misgina Tsighe Hagos, Shri Kant, Surayya Ado Bala \\
Research and Technology Development Center\\
Sharda University\\
Greater Noida, India \\ 
\texttt{\{tsighemisgina,shrikant.ojha,surayyaadob\}@gmail.com}
}

\iclrfinalcopy 
\begin{document}

\maketitle

\begin{abstract}
Early diagnosis of diabetic retinopathy for treatment of the disease has been failing to reach diabetic people living in rural areas. Shortage of trained ophthalmologists, limited availability of healthcare centers, and expensiveness of diagnostic equipment are among the reasons. Although many deep learning-based automatic diagnosis of diabetic retinopathy techniques have been implemented in the literature, these methods still fail to provide a point-of-care diagnosis. This raises the need for an independent diagnostic of diabetic retinopathy that can be used by a non-expert.  Recently the usage of smartphones has been increasing across the world. Automated diagnoses of diabetic retinopathy can be deployed on smartphones in order to provide an instant diagnosis to diabetic people residing in remote areas. In this paper, inception based convolutional neural network and binary decision tree-based ensemble of classifiers have been proposed and implemented to detect and classify diabetic retinopathy. The proposed method was further imported into a smartphone application for mobile-based classification, which provides an offline and automatic system for diagnosis of diabetic retinopathy.
\end{abstract}

\section{Introduction}

\subsection{Diabetic Retinopathy Diagnosis}

Diabetic Retinopathy (DR) is a retinal complication caused when retinal blood vessels are damaged by diabetes. Signs of DR start with microaneurysms, which are small red spots that appear when there is a blood escape from retinal blood vessels. If microaneurysms are not treated early walls of capillaries may get broken which form hemorrhages. Exudates may appear on the retina if treatment of the disease is delayed, and this can lead to permanent vision loss or vision impairment.


DR diagnosis is clinically performed by ophthalmologists with the help of high-end fundus images capturing devices. In order to capture retinal images, different imaging techniques, such as optical coherence, tomography, and fundus photography, have been used \citep{salz2015imaging}. All of these techniques come with the challenge of expensive design, deployment, and usage costs. Trained professionals are needed to use these techniques. In addition to the trained professional need of these techniques, an ophthalmologist or more are required to study and diagnose a fundus image that is captured by the imaging methodology. An ophthalmologist usually requires two to seven days for retinal image diagnosis. Diabetes patients that reside in rural and remote areas usually suffer from delayed diagnosis and treatment of DR because of the expensive deployment of diagnosing equipment and shortage of ophthalmologists and health care centers.

Limited access to point-of-care diagnostic services was identified as one of the barriers of medical diagnosis in rural areas \citep{huaynate2015diagnostics}. \citet{foster2005impact} put four strategies to fight the challenges the diagnoses process of DR faces in order to implement treatment for preventable blindness; (1) Creating academic, public and governmental awareness of the effects of blindness and visual loss, and the fact that 75\verb|%| of diseases that cause blindness are preventable; (2) Automating and mobilizing existing techniques and methods; (3) Implementing district-specific and country-specific prioritizing strategies of diagnosing and treatment resources for a productive process; (4) Providing comprehensive, maintainable and fair diagnosis services of visual diseases at district level, which includes staff  training, distributing diagnosis and treatment resources, and infrastructure, such as health care center buildings.

A timely and accurate automatic diagnosis of DR could enable diabetic patients to get treatment for preventable visual diseases, thereby avoiding permanent vision loss or impairment. Point-of-care DR diagnostic service, which aims to diagnose DR instantly at a patient’s place, is achieved in this paper using a smartphone for data collection and trained model for detection.

\subsection{Annotated Training Data}

One of the main issues with incorporating deep learning in medical image analyses is the shortage of available annotated training dataset \citep{miotto2017deep}\citep{razzak2018deep}. Transfer learning techniques have gained wider acceptance because of the unavailability of enough annotated training data in the design and training of deep convolutional neural network models \citep{erhan2010does}\citep{litjens2017survey}. In \citet{altaf2019going}, annotated training data insufficiency was identified as the main challenge of applying deep learning models in the healthcare automation industry. Furthermore, \citet{altaf2019going} recommended that methods that exploit deep learning using reduced data need to be devised and implemented. We have combined inception module based convolutional neural networks with a binary tree-based ensemble of classifiers to increase the performance of our proposed neural network model with limited training data.

This paper is structured into five sections. Section \ref{related_work} presents a review of related works. The proposed methodology and implementation are discussed in Section \ref{method}. Results and discussions are presented in Section \ref{results}.  Lastly, we conclude in Section \ref{conclusion}.

\section{Related Work}
\label{related_work}

Although the literature of automated DR detection and classification contains various works, the methods employed can be generally classified into two, which are feature extraction based and image-level based classifications. In order to provide an onsite diagnosis of DR, mobile detection works have also been proposed and implemented. Deep learning models such as Convolutional Neural Networks (CNN) are usually used to classify fundus images.  A fundus images dataset that is uploaded to the Kaggle website for DR detection competition \citep{kaggle}  has been extensively used in the image level DR classification literature. End-to-end approach has been used to train a deep learning model from scratch.  In order to avoid the cost of training, transfer learning has also been used.

In multi-class classification, DR has been classified into one of five stages, which are normal, mild, moderate, severe, and proliferative stages, as proposed by \citet{wilkinson2003proposed}. In the following subsections, different works that have been performed in detecting DR will be reviewed.

\subsection{Feature Extraction Based Classification of Diabetic Retinopathy}

Lesions of DR such as macular edema, exudates, microaneurysms, and hemorrhages have been automatically identified, segmented and detected in order to classify DR. End-to-end and transfer learning of Convolutional Neural Networks (CNN) and Fully Convolved Residual Networks (FCRN) have been employed for classifying the extracted lesions. Table \ref{feature-based} (See Appendix A) presents a summary of lesion detection based approaches for detecting DR.

\subsection{Image Level Classification of Diabetic Retinopathy}

In image level-based classification of DR from fundus images, there is no need to segment and extract lesions. The detection is performed on the whole fundus image. Table \ref{image-based} (See Appendix A) summarizes the image level DR detection literature.

\subsection{Mobile Classification of Diabetic Retinopathy}

One of the strategies suggested by \citet{foster2005impact}  for prevention of treatable blindness was the distribution of maintainable and unbiased eye care services at the district level. Tele-retina has been implemented to capture retinal images by non-experts from remote areas and transferring to the ophthalmic center for DR diagnosis \citep{cavallerano2004telehealth}\citep{bursell2012telemedicine}\citep{coronado2014diagnostic}. Tele retina needs a strong network connection between patients and health care centers. Teleophthalmology, which includes analysis of collected images, can be implemented to provide DR diagnosis remotely \citep{mohammadpour2017smartphones}.

As is put by \citet{foster2005impact}, in order to automate and mobilize existing techniques and methods, a smartphone device with mobile retinal image capturing cameras could be used to diagnose DR in remote and rural areas. This can solve the cost and time challenges of the traditional diagnosis of DR.

One of the main challenges of deploying DR automatic diagnosis on a smartphone is the quality of the retinal images captured and the limited processing power and memory inside. A speedy, easy to use, and cheap retinal imaging device could be used as an add-on with a smartphone to collect fundus images of patients as in \citet{kim2018smartphone}.

Smartphone based automatic diagnosis of DR can be implemented in two ways. The first way is an internet-based diagnosis, and the second one is a stand-alone independent smartphone application software-based diagnosis. In an internet-based diagnosis, the smartphone is only used to capture a fundus image of a patient with the help of an add-on retinal camera. After a fundus image is captured, it is sent to a professional ophthalmologist for diagnosis, and results would be sent back over the internet. This would require the user to have a stable internet connection for sending the fundus images and receiving back the results of the diagnosis. The challenge of low internet coverage can be solved using an independent smartphone application that is able to process a captured fundus image and automatically diagnose the stage of DR without any network connection.

Table \ref{mobile-based} (See Appendix A) summarized the review of mobile-based classification services developed for DR classification. \citet{rajalakshmi2018automated} resulted in the highest performance by using a cloud-based diagnostic service while the fundus image is captured with a smartphone. 

\section{Proposed Methodology}
\label{method}

\begin{figure}[h]
\centering
\subfigure[Development process]{\label{proposed_system_fig}\includegraphics[width=0.4\linewidth]{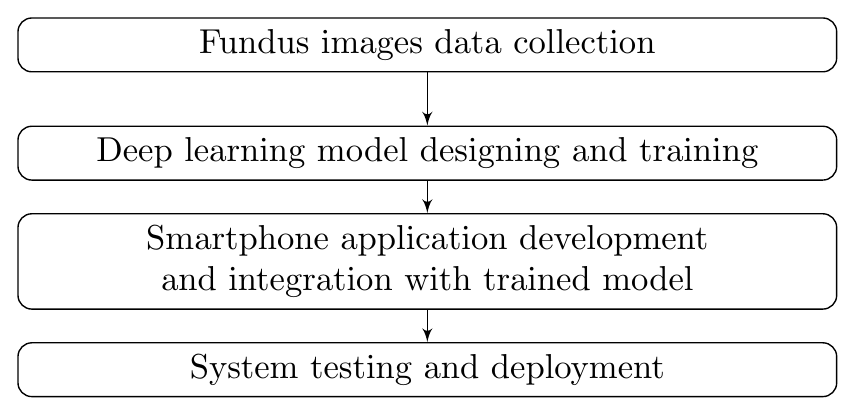}}
\subfigure[Binary tree based ensemble of binary classifiers]{\label{binary_tree}\includegraphics[width=0.5\linewidth]{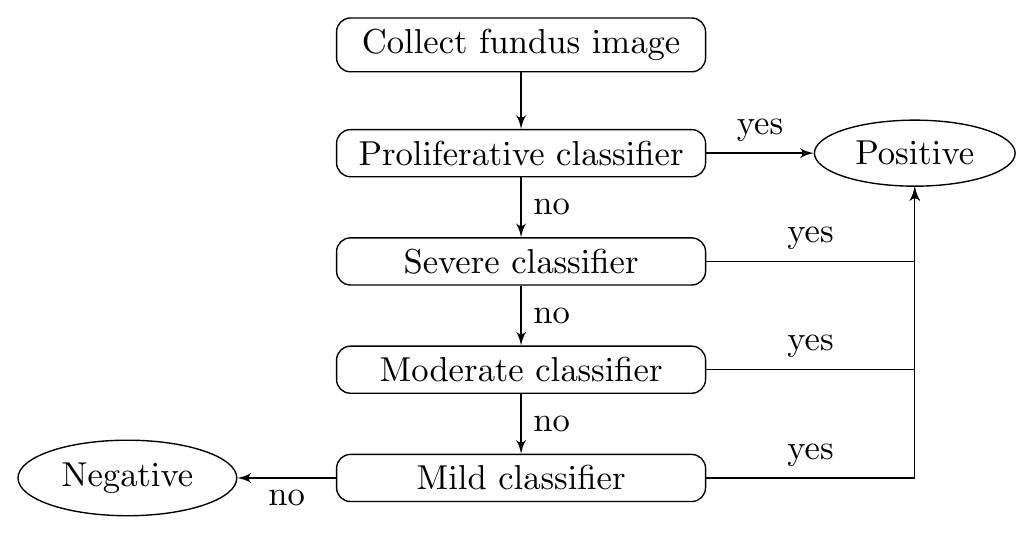}}
\label{proposed_with_tree}
\caption{Development process and binary tree based ensemble of binary models.}
\end{figure}

In order to provide an easy to use and independent smartphone based diagnoses of DR, a deep learning model needs to be trained first. After training, the model needs to be deployed in a smartphone application. For clarification, Figure \ref{proposed_system_fig} presents the flow of the proposed development process.

\subsection{Deep Learning Model Training}

For training a deep learning model, a dataset of fundus images that are uploaded to Kaggle website \citep{kaggle} has been used. The dataset contains fundus images labeled into five stages of DR. Inspired by the inception module presented in \citet{szegedy2015going} inception based convolutional neural network was designed and implemented. We further discuss image pre- processing, inception based network, and binary decision tree-based ensemble of classifiers.

\subsubsection{Image Pre-processing}

\begin{figure}[h]
\centering
\subfigure[Original image]{\label{img_orig}\includegraphics[width=1.65cm, height=1.45cm]{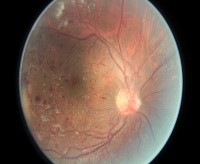}}
\hspace{1cm}
\subfigure[Intermediate output]{\label{img_processed_1}\includegraphics[width=1.60cm, height=1.45cm]{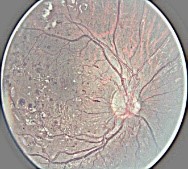}}
\hspace{1cm}
\subfigure[Preprocessed fundus image. ]{\label{img_processed_2}\includegraphics[width=1.45cm, height=1.45cm]{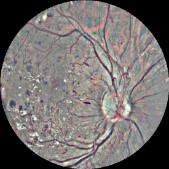}}
\caption{Image pre-processing on a sample fundus image. A sample input fundus image is shown in (a). (b) Cropped and local average subtracted from the original image. (c) 90\% of Figure \ref{img_processed_1}.}
\label{preprocessing}
\end{figure}

For model training, 200 images were selected from each stage of DR. In order to exploit high performance from a deep learning model and smaller training dataset; image pre-processing was performed by adopting \citet{graham2015kaggle}'s method. The fundus images were cropped to remove unnecessary black background pixels. The local average was subtracted from the cropped fundus image. Cropping of the images’ 10\% was repeated to remove white rounding noise. Figure \ref{img_orig}, \ref{img_processed_1} and \ref{img_processed_2} show the implemented preprocessing on a sample fundus image.

\subsubsection{Inception based Network}

\begin{figure}[h]
\centering
\subfigure[Model framework]{\label{framework}\includegraphics[width=0.4\linewidth]{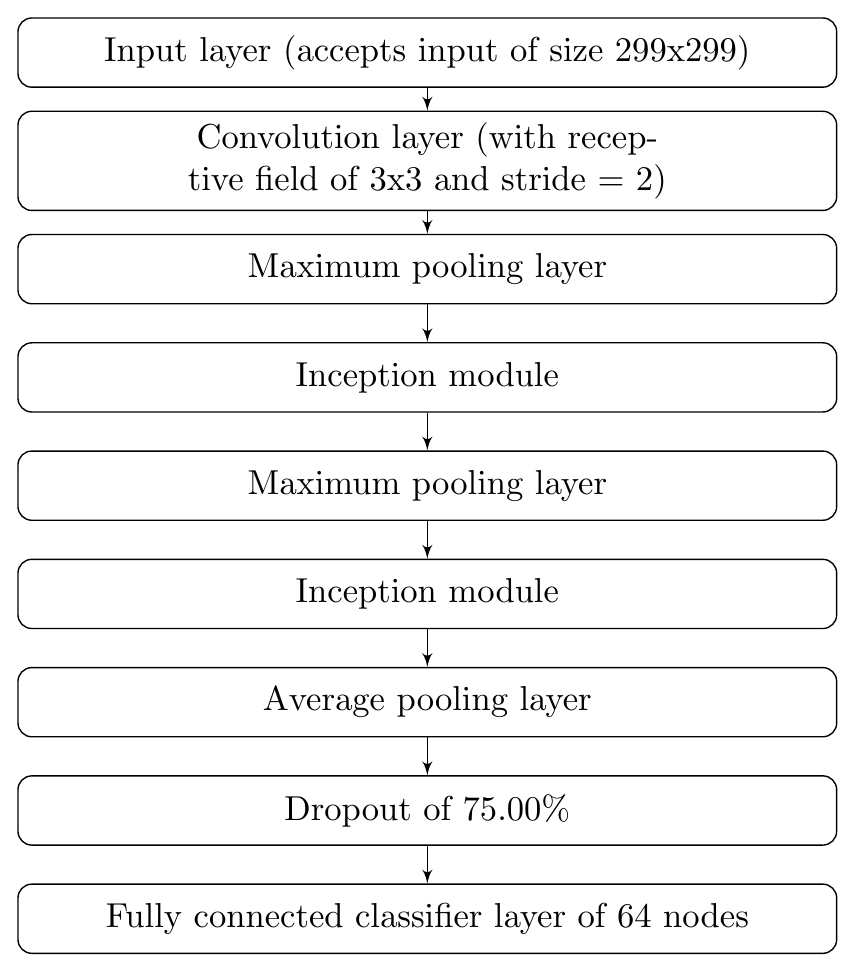}}
\subfigure[Inception module]{\label{inception}\includegraphics[width=0.4\linewidth]{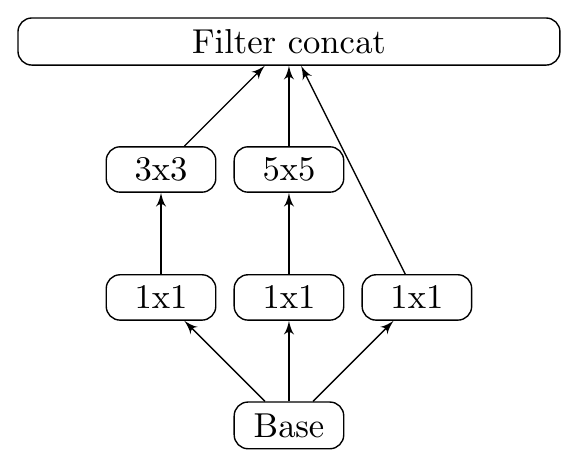}}
\label{side}
\caption{Framework of the designed model and inception module.}
\end{figure}

After the automatic diagnosis is performed, the designed deep learning model should be smartphone adaptable in order to be imported into a mobile application and fit to the memory limit of smartphones. For this reason, an inception module based convolutional neural network was designed. The inception module’s architecture is presented in Figure \ref{inception}. It was designed based on the inception module presented in \citet{szegedy2015going}.

The framework of our proposed inception based convolutional neural network is illustrated in Figure \ref{framework}. The network was trained using Stochastic Gradient Decent (SGD) with an ascending learning rate of 0.0001.

\subsubsection{Binary Decision Tree-based Ensemble of Classifiers}

In order to increase the performance of our proposed inception based convolutional neural network, a binary tree-based ensemble of four binary classifiers were used. Four binary classifiers were trained to classify between normal and mild, normal and moderate, normal and severe, and normal and proliferative stages of DR. The mild, moderate, severe and proliferative stages of the disease are considered to be unhealthy states of the retina. Figure \ref{binary_tree} presents our proposed binary decision tree-based ensemble of classifiers.

\subsection{Smartphone Application Development}

A standalone smartphone application software was designed to import the trained binary deep learning networks and implement a binary decision tree-based ensemble to classify fundus images into healthy and unhealthy. The smartphone device’s operating system used was ColorOS version 6.0, which is a variety of Android. Size of the trained binary classifying models was 5757 kilobytes.

A retinal image is proposed to be captured using add on camera or from the smartphone’s gallery by tapping on the “GET” button as is shown in Figure A.\ref{app_interface}. And the label of the imported retinal image is predicted by tapping on the “CLASSIFY” button.

\section{Results and Discussion}
\label{results}

In this section, we compare the performance of our trained model with  similar works, which have classified DR into binary classes using the Kaggle dataset \citep{kaggle}, as shown in Table \ref{results_table}.

\begin{table}[h] 
\caption{Comparison of parameters and results}
\label{results_table}
  \centering
  \renewcommand{\arraystretch}{1.2}
  \begin{tabular}{p{2cm}p{2cm}p{2cm}p{2cm}p{2cm}}
    \multicolumn{2}{l}{\textbf{POINTS OF COMPARISON}}& \textbf{\citet{mohammadian2017comparative}}& \textbf{\citet{hagos2019transfer}}& 
    \textbf{PROPOSED METHOD} \\
    \hline
    \textbf{Training parameters}	&Training dataset size&	35126&	2500&	1000\\
    &	Optimizer&	ADAM&	SGD	& SGD\\ 
    &	Learning rate& -& Ascending 0.0001&	Ascending 0.0001\\
    &	Loss function	&-&	Cosine loss &	Cosine loss  \\
    &	Data augmentation used?	&Yes&No&	No\\
    \hline
    \textbf{Results}	&	Accuracy	&87.12\%&	90.90\%&	99.86\\
    &	Sensitivity(\%)	&-&	-&	99.25  \\
    &	Specificity(\%)	&-&	-&	99.60 \\
    &	Loss(\%)	&-&	3.94&	-
  \end{tabular}
\end{table}

From Table \ref{results_table}, our proposed model achieved the best performance with 99.86\% accuracy, 99.25\% sensitivity and 99.60\% specificity followed by the work of \citet{hagos2019transfer} and \citet{mohammadian2017comparative}. From this analysis, it shows that our proposed method with limited training data, and without data augmentation performs better than other previous methods. Figure A.\ref{app_result} shows our developed smartphone application's classification output on a sample fundus image.

\section{Conclusion}
\label{conclusion}

In this paper, inception based convolutional neural network, binary decision tree-based ensemble of classifiers, and android smartphone application development have been combined in order to provide a standalone smartphone based automatic diagnosis of DR with an accuracy of 99.86\%. Using our implemented system, diabetic people residing in remote and rural areas with less coverage of healthcare centers and shortage of professional ophthalmologists only need an Android smartphone, and a portable fundus camera add on for an instant DR diagnosis. The methods followed in this paper can lead the way in providing a non-expert usable point-of-care diagnostic services for DR in specific and other medical image classification tasks in general. At this stage, the developed application only diagnoses images from a phone's gallery. For future work, we plan to perform a multi-class classification of DR, and incorporate a handheld fundus camera to start deployment. 

\bibliography{DL_with_app}
\bibliographystyle{iclr2020_conference}

\appendix
\section{Appendix}

\renewcommand\thefigure{\thesection.\arabic{figure}}
\renewcommand\thetable{\thesection.\arabic{table}}
\setcounter{figure}{0} 
\setcounter{table}{0} 

\begin{table}[b]
\caption{Summary of Feature Extraction Based Diabetic Retinopathy Classification}
\label{feature-based}
  \centering
  \renewcommand{\arraystretch}{1.2}
  \begin{tabular}{p{2cm}p{2cm}p{2cm}p{2cm}p{1cm}p{1cm}p{1cm}}
    \multirow{2}{2cm}{\textbf{AUTHORS}}& \multirow{2}{2cm}{\textbf{METHODS}}& \multirow{2}{2cm}{\textbf{TRAINING}}& \multirow{2}{2cm}{\textbf{LESIONS}}& \multicolumn{3}{l}{\textbf{PERFORMANCE}} \\
    \cline{5-7}&&&&\textbf{SN(\%)} & \textbf{SP(\%)} & \textbf{ACC(\%)}\\
    \hline
    \citet{perdomo2016novel} & CNN & Transfer learning & Macula edema & 56.50 & 92.80 & 77.00 \\
    \citet{prentavsic2016detection} & CNN with 11 layers & End-to-end & Exudates & 78.00 & - & - \\ 
    \citet{abramoff2016improved} & CNN & End-to-end & Macula edema & 100.00 & - & -  \\
    \citet{perdomo2017convolutional} & LeNet CNN & Transfer learning & Exudates & 99.80	& 99.60 &	99.60  \\
    \citet{mo2018exudate} & FCRN & End-to-end & Macula edema & 92.55 &	- &	-  \\
    \multirow{2}{2cm}{\citet{gondal2017weakly} } &
    \multirow{2}{2cm}{Octree based CNN and Transfer learning} & Transfer learning & Microaneurysms & 52.00	&-&	-\\
    &&& Haemorrhages&	91.00	&-&	-
  \end{tabular}
\end{table}

\begin{table}[h]
\caption{Summary of Image-Level Diabetic Retinopathy Classification}
\label{image-based}
  \centering
  \renewcommand{\arraystretch}{1.2}
  \begin{tabular}{p{2cm}p{2cm}p{2cm}p{1cm}p{1cm}p{1cm}p{1cm}}
    \multirow{2}{2cm}{\textbf{AUTHORS}}& \multirow{2}{2cm}{\textbf{METHODS}}& \multirow{2}{2cm}{\textbf{TRAINING}}& \multicolumn{4}{l}{\textbf{PERFORMANCE}} \\
    \cline{4-7}&&&\textbf{SN(\%)} & \textbf{SP(\%)} & \textbf{ACC(\%)}& \textbf{AUC}\\
    \hline
    \citet{gulshan2016development} & Inception-V3 CNN &	Transfer learning&	90.30&	90.00&	-&	0.99 \\
    \citet{colas2016deep} & CNN	&End-to-end	& 96.20	& 66.60 &	-	& 0.95 \\ 
    \citet{pratt2016convolutional} & CNN with 13 layers &	End-to-end &	95.00 &	30.00	& 75.00	& - \\
    \citet{mohammadian2017comparative} & Inception-V3 &	Transfer learning &	- &	-	& 87.12	& -  \\
    \citet{quellec2017deep} & AlexNet CNN &	Transfer learning &	-	& -	& -	&0.95  \\
    \citet{costa2017convolutional} & CNN&	End-to-end&	78.00&	-	&-	& 0.97  \\
    \citet{abramoff2016improved} & CNN	&End-to-end	&96.80	&87.00&	-&	0.98 \\
    \citet{gargeya2017automated} & ResNet&	End-to-end&	94.00&	98.00&	-&	-  \\
    \citet{ting2017development} & CNN&	End-to-end&	90.5&	91.60	&-&	0.94  \\
    \citet{Dutta2018} & Deep neural network	&End-to-end&	-	&-&	86.30&	- \\
    \citet{Mateen2018} & VGGNet-19 with SVM	&End-to-end	&-&	-	&98.34&	-\\
    \citet{Mansour2018} & AlexNet with SVM classifier&	Transfer learning&	100.00&	93.00&	97.93&	-  \\
    \citet{orlando2018ensemble} & CNN	&End-to-end&	97.21&	50.00&	-&	0.93  \\
    \citet{Adly2019} & VggNet &	End-to-end &	81.80&	89.30&	83.20&	-  \\
    \citet{nagasawa2019accuracy} & CNN	&End-to-end&	94.70&	97.20	&-&	0.97  \\
    \citet{verbraak2019diagnostic} & Detection Validation of a device	&End-to-end&
	100.00	&-&	-&	-
 \\
    \citet{gonzalez2019evaluation} & Performance validation&	End-to-end&
	92.00	&92.10	&-	&0.97
  \\
    \citet{hagos2019transfer} & Inception-V3 network&	Transfer learning with fine-tuning&	-	&-&	90.90&	-  \\
  \end{tabular}
\end{table} 

\makeatletter
\setlength{\@fptop}{0pt}
\makeatother

\begin{table}[ht!]
\caption{Summary of mobile based diabetic retinopathy classification}
\label{mobile-based}
  \centering
  \renewcommand{\arraystretch}{1.2}
  \begin{tabular}{p{2cm}p{2cm}p{2cm}p{1cm}p{1cm}p{1.7cm}p{1cm}}
    \multirow{2}{2cm}{\textbf{AUTHORS}}& \multirow{2}{2cm}{\textbf{TECHNIQUES}}& \multirow{2}{2cm}{\textbf{CAMERA}}& \multicolumn{4}{l}{\textbf{PERFORMANCE(\%)}} \\
    \cline{4-7}&&&\textbf{SN} & \textbf{SP} & \textbf{PRECISION}&\textbf{ACC}\\
    \hline
    \citet{prasanna2013decision} & Feature extraction based classification	&ophthalmoscope&	-&	-&	-&	- \\
    \citet{xu2016smartphone} & Vessel segmentation &	-&	-&	-	&-&	93.30 \\ 
    \citet{kashyap2017mobile} & DWT based classification	&	LED with a condensing lens &	57.00&
	-&	63.00&	-\\
    \citet{rajalakshmi2018automated} & Cloud-based diagnostic	&	Smartphone	&99.10&	80.40&	-&	-  \\
    \citet{jamil2018smart} & Performance validation of mobile diagnosis	&	20 dioptre condensing lens with a phone camera	&- &-&	-&	-\\
    \citet{kashyap2019color} & Histogram comparison	&	LED with a condensing lens	&53.00&	-&	62.00&	-
  \end{tabular}
\end{table}

\begin{figure}[h]
\centering
\subfigure[User interface]{\label{app_interface}\includegraphics[width=0.4\linewidth]{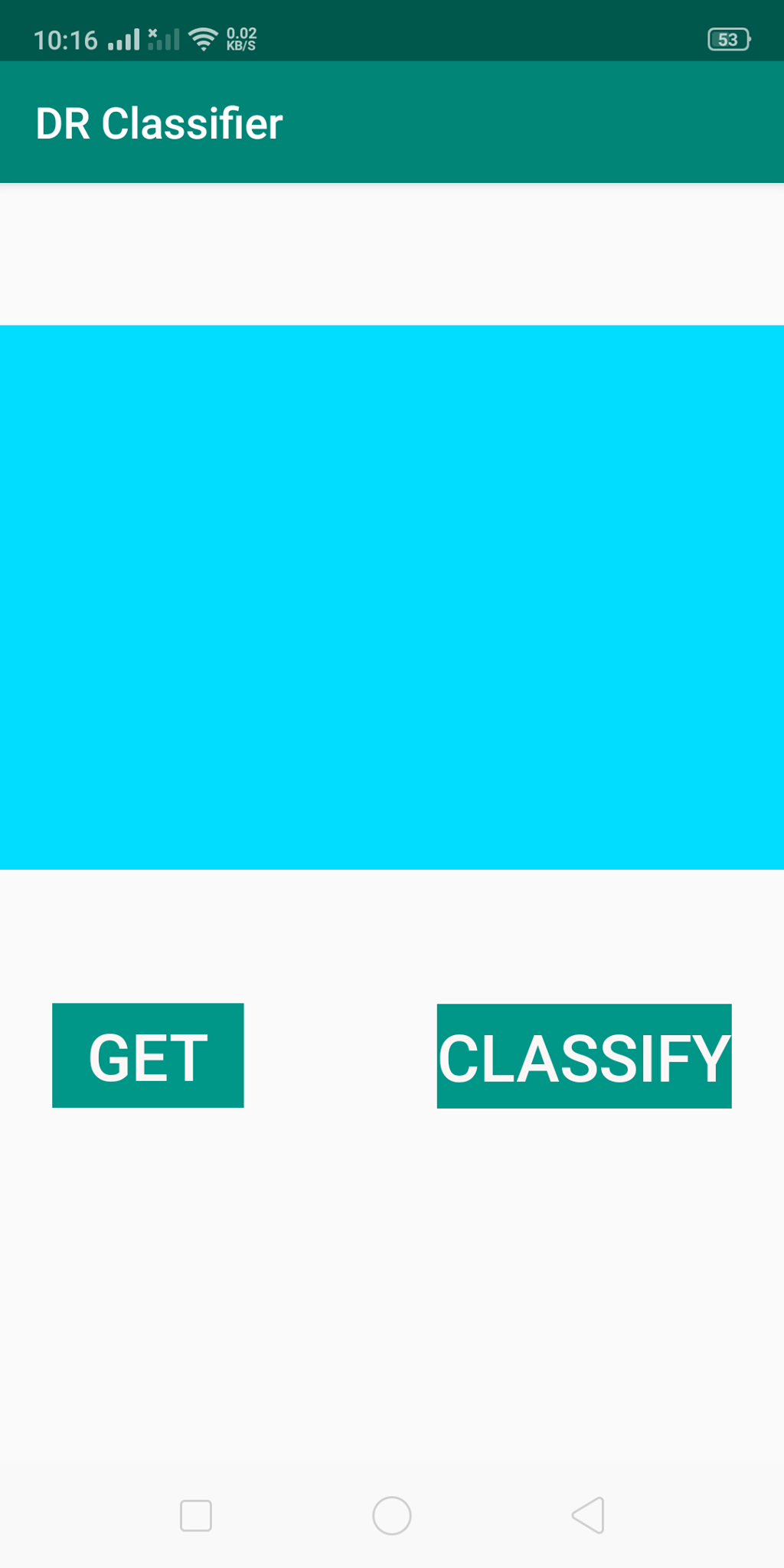}}
\hspace{1cm}
\subfigure[Classifier result]{\label{app_result}\includegraphics[width=0.4\linewidth]{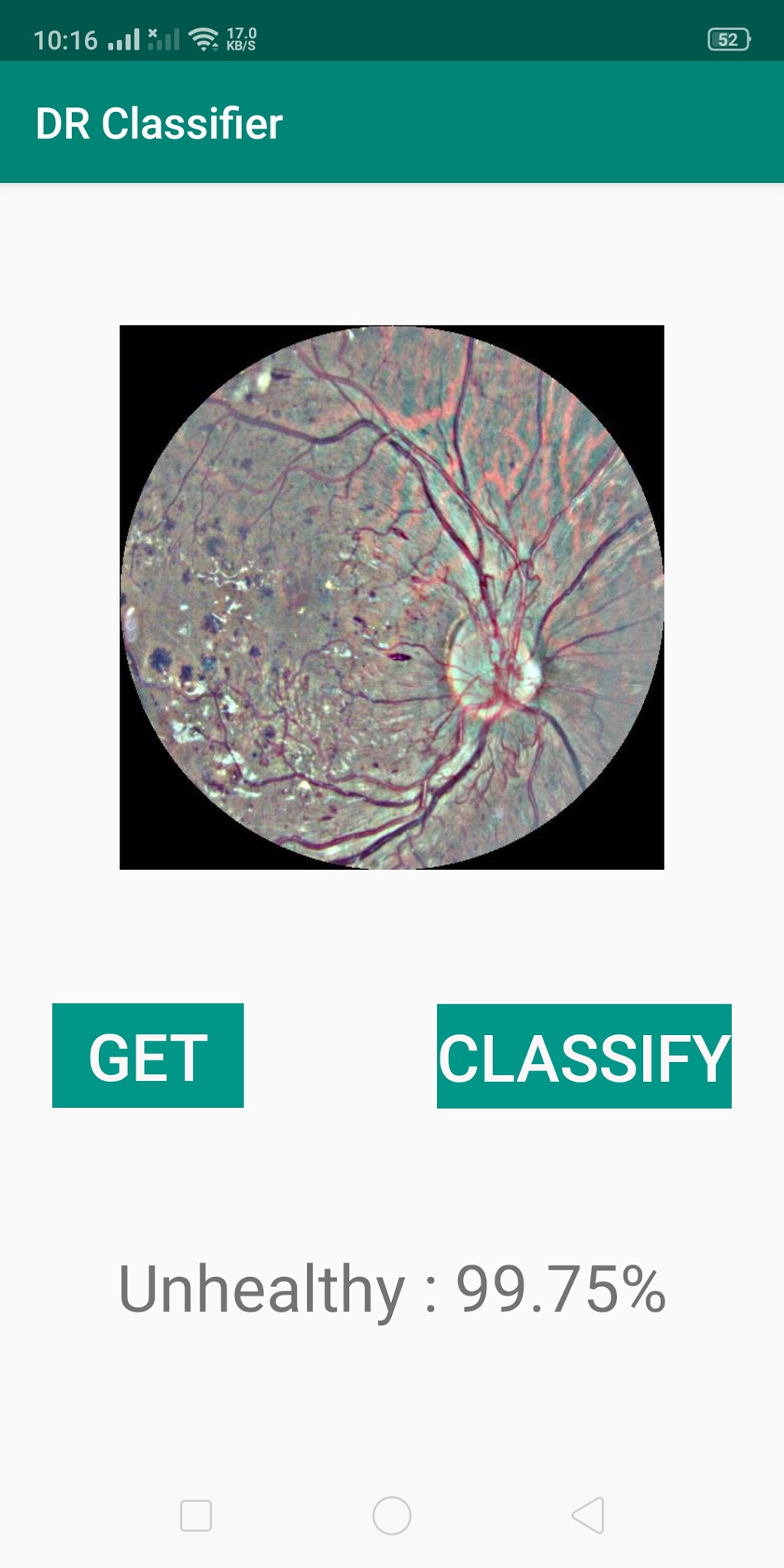}}
\caption{Smartphone based diagnosis.}
\label{app}
\end{figure}

\end{document}